# ASTRI for the Cherenkov Telescope Array

**Maria Concetta Maccarone**[1]

*Istituto Nazionale di Astrofisica, INAF – IASF Palermo, Via Ugo La Malfa 153, I-90146 Palermo, Italy*
*E-mail:* `Cettina.Maccarone@iasf-palermo.inaf.it`

**for the CTA ASTRI Project**
http://www.cta-observatory.org
http://www.brera.inaf.it/astri

The Cherenkov Telescope Array (CTA) will be the largest ground-based observatory operating in the very-high-energy gamma-ray (20 GeV – 300 TeV) range. It will be based on more than one hundred telescopes, located at two sites (in the northern and southern hemispheres). The energy coverage, in the southern CTA array, will extend up to hundreds of TeV thanks to 70 small size telescopes (SST), with their primary mirrors of about 4 meters in diameter and large field of view of the order of 9 degrees. It is proposed that one of the first sets of precursors for the CTA SSTs array will be represented by (at least) nine ASTRI telescopes. The prototype of such telescopes, named ASTRI SST-2M, is installed in Italy. It is currently completing the overall commissioning before entering the science verification phase that will performed observing bright TeV sources as Crab Nebula, Mrk421 and Mrk 501. This science verification phase will cross-check the prototype performance with the predictions of detailed Monte Carlo simulations.

ASTRI telescopes are characterized by a dual-mirror optical design based on the Schwarzschild-Couder (SC) configuration. The focal-plane camera is curved in order to fit the ideal prescription for the SC design and the sensors are small size silicon photomultipliers read-out by a fast front-end electronics. The telescope prototype installed in Italy, has been developed by the Italian National Institute for Astrophysics, INAF, following an end-to-end approach that comprises all aspects from the design, construction and implementation of the entire hardware and software system to the final scientific products. In this respect, the internal and external calibration systems, control/acquisition hardware and software, data reduction and analysis software, and data archiving system are being realized in addition to the telescope structure and camera. All parts of the system have been designed to comply with the CTA requirements. A collaborative effort, addressed to the implementation of the first ASTRI telescopes for the CTA southern site, is now on-going led by INAF with the Universidade de Sao Paulo (Brazil), the North-West University (South Africa) and the Italian National Institute for Nuclear Physics, INFN. In this contribution we will describe the main features of the ASTRI telescopes, the latest news from their prototype, its performance and the expectations of the first set of ASTRI telescopes proposed for CTA.

---

[1]Speaker

## 1. Introduction

The Cherenkov Telescope Array (CTA) [1, 2] is the next generation ground-based observatory for gamma-ray astronomy at very high energies. With more than one hundred Cherenkov telescopes located in two sites, one for each hemisphere, assuring full sky coverage, CTA will detect very high energy radiation with unprecedented accuracy and sensitivity better than current Image Atmospheric Cherenkov Technique (IACT) experiments. Three different classes of telescopes (LST, MST, SST, that is large, medium and small, so defined with respect to their collecting mirror size) will have different field-of-views and are designed to access different energy regimes starting from 20 GeV. The CTA energy range between a few TeV and 300 TeV is envisaged to be covered by 70 SSTs spread out over several square kilometers in the CTA southern site. Three types of SSTs are foreseen, one with single mirror and two with double mirror optics [2, 3]: the ASTRI telescopes are one of the implementations with double-mirror design. In particular, they constitute the main component of the ASTRI project, led by the Italian National Institute for Astrophysics (INAF).

The ASTRI project, acronym for "Astrofisica con Specchi a Tecnologia Replicante Italiana", started in 2011 as a 'Flagship Project' with funds provided by the Italian Ministry of Education, University and Research and specifically assigned to INAF for CTA. The first goal has been the design, development and deployment of an end-to-end prototype for the CTA small size class of telescopes in a dual-mirror configuration (2M). The prototype, named ASTRI SST-2M, is installed in Italy. The ASTRI project continues to be led by INAF and supported by the Italian government with specific funds. Moreover it has involved a wider community inside CTA and it is now a collaborative international CTA effort carried on by Italian, Brazilian and South-African Institutes. Starting from 2019, the ASTRI project aims of deploying a first set of nine ASTRI telescopes [4] so contributing to the implementation of the initial partial southern array of CTA. Final aim is the installation of at least 35 ASTRI telescopes out of the 70 small-sized telescopes of the CTA.

The main features and performance of the ASTRI telescopes will be described in the following sections. Technical details can be found in the citations along the text and references therein.

## 2. ASTRI: the telescopes

The ASTRI telescopes present several innovative technological solutions for the detection of atmospheric Cherenkov light. From a general point of view, like every IACT telescope, they have a mount that allow them to rapidly point towards targets and are comprised of a segmented collecting mirror system to reflect the Cherenkov light to a high-speed camera that can digitize and record the image of the shower.

What are then the innovations? First of all, the optical system that implements, for the first time, the double mirror Schwarzschild-Couder (SC) configuration [5, 6]. This kind of SC telescope has never been realized until CTA, mainly due to technological difficulties. However, recent advances in technology (in particular for the realization of the aspherical mirrors) have made the implementation of this design practicable for the observation of atmospheric Cherenkov light. The 4.3 m diameter primary mirror is composed of an array of hexagonal tiles, each tile having been manufactured by means of the cold slumping replication process in which a thin sheet of glass is formed at room temperature over a mould [7]. The ASTRI secondary mirror, 1.8 m diameter, is a monolithic hemispherical thick glass shell thermally bent to 2.2 m radius of curvature. The telescope equivalent focal length is 2.15 m, f/0.5, and the system covers a full field-of-view (FoV) of more than 10° [8].

The double mirror SC configuration provides a much better Point Spread Function (PSF) over large field of view compared to a single mirror. It allows better correction of aberrations at large off-axis angles and facilitates the construction of compact telescopes. Moreover, this configuration results in a small plate scale (37.5 mm/deg), leading to a camera of compact dimension. Having typical lateral dimension of few millimeters, the silicon photomultipliers (SiPM) sensors are the perfect solution to realize such a compact Cherenkov camera with large FoV. Assembled without any light collection system onto the ASTRI curved focal plane located between the two mirrors, the SiPMs act as the camera pixels. In the ASTRI camera the SiPM sensors are organized in 37 Photon Detection Modules (PDM) of 8×8 pixels, each of them with a sky-projected angular size of 0.19°, matching the angular resolution of the optical system. This configuration yields a PSF, defined as the 80% of the light collected from a point like source, contained in less than one camera pixel.

The SiPM sensors that have been chosen for all the CTA small size telescopes exhibit very fast response and excellent single photoelectron resolution. They need a properly tailored electronics devoted to directly interface the SiPM sensors, detecting the fast pulses produced by the Cherenkov flashes, managing the trigger generation, the digital conversion of the signals and the transmission of the data to the camera server external to the telescope. Differently from other CTA telescopes that use the standard sampling technique, the ASTRI camera electronics is based on a custom peak-detector operation mode to acquire the SiPM pulses [9, 10]. The Cherenkov Imaging Telescope Integrated Read Out Chip (CITIROC), with its signal shaper and peak detector customized for ASTRI, represents an innovative technical solution and provides high efficiency pixel by pixel trigger capability and very fast camera pixel read out [11].

The ASTRI camera trigger is a topological one, activated when a given number of contiguous pixels within a PDM presents a signal above a given photo-electron threshold. Both the number of contiguous pixels required for a trigger and the signal threshold can be manually set or automatically adjusted, depending on the level of night sky background. Data registered by the ASTRI camera at the occurrence of a trigger condition (scientific raw event) or when required (calibration or housekeeping data) are then sent to the camera server in which the camera data acquisition software is installed [12].

To complete this general overview, it is worth to mention that the ASTRI focal plane is covered by a protective window and the entire camera, with its electronics and all ancillary devices, is closed in a basket whose top is open thanks to a lid system. A novel inner fiber-optic equipment for relative calibration [9, 13, 14], and a GPS receiver used for time synchronization and providing tag-time for triggered events, are part of the camera ancillary devices. A sketch of the ASTRI camera is shown in Fig.1 (right).

### 2.1 The ASTRI SST-2M telescope prototype installed in Italy

The ASTRI SST-2M prototype has been developed following an end-to-end approach that comprises all of the work that should be done to achieve the goal, i.e. the validation of the adopted technological solutions and then the final scientific products. This approach includes development, deployment, operation and control of structure, mirrors, camera, inner calibration system, software and hardware for control and acquisition, data reduction and analysis software, data archiving system and the external equipment for monitoring and calibration purposes.

The prototype is shown in Fig.1 (left). It is installed in Italy at the INAF "M.C. Fracastoro" observing station located in Serra La Nave (Mt. Etna, Sicily), 1740 m a.s.l. [15]. In the station different telescopes are operated. Moreover the station is equipped with various instrumentation devoted to the monitoring of meteorological and environmental conditions [16]. Finally a dedicated control room and a data centre have been developed as part of the ASTRI project [17].

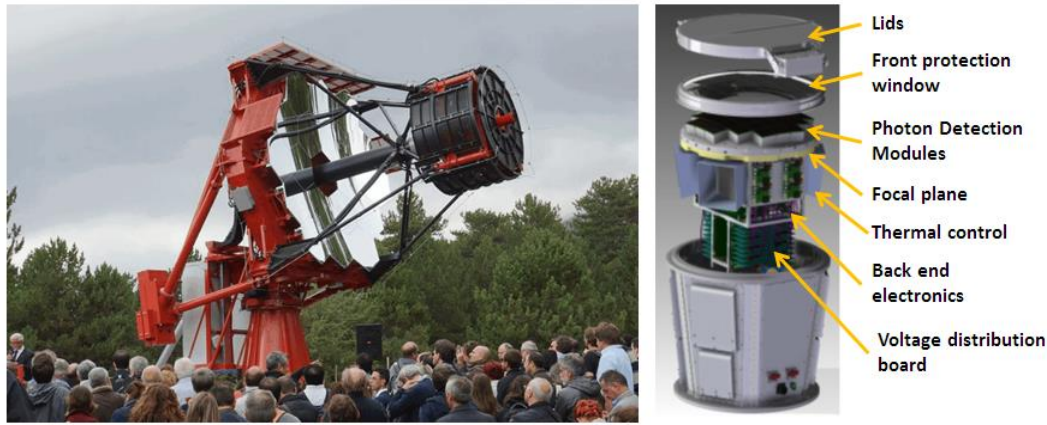

Fig.1 –The ASTRI SST-2M telescope prototype installed at the Serra La Nave (left) and exploded sketch of its camera (right).

The ASTRI SST-2M telescope prototype, whose structure was inaugurated in September 2014, is currently completing the overall commissioning phase being its camera installed at the focal plane. Several tests have been already conducted and the necessary control/acquisition systems optimized. Among them, the telescope control system to move, point and track the rigid altitude-azimuthal telescope structure [18, 19] and the camera system to manage and control all its components, from the trigger to the read-out electronics until the ancillary devices and the lid motors [20]. The graphical user interfaces, that provide full access to the capabilities offered by the telescope hardware subsystems for testing and maintenance [21], have been optimized as well as the monitoring and alarms system for the necessary information and communication (ICT) infrastructures [17]. Functionality test have been performed on the camera server and data acquisition system aimed at the acquisition, pre-processing, storage and monitoring of the camera data, including muon ring images candidate for calibration purposes [12, 22]. The system includes a Quick Look, with event viewer, able to perform simple real-time analysis. The camera data acquisition system interacts with the data handling subsystem in order to transfer the files, after proper format conversion, to the data archive and scientific analysis components [23-25]. Eventually, an absolute calibration strategy has been defined making use of auxiliary equipment external to the prototype, namely the Illuminator and UVscope. The Illuminator is a portable ground-based device, remotely controlled, designed to uniformly illuminate, from a certain distance, the telescope aperture with a pulsed or continuous reference photon flux whose absolute intensity is monitored by a NIST-calibrated photodiode. Using different illumination features (wavelength, intensity, pulse length) as well as changing the telescope pointing towards the Illuminator, several calibration purposes can be accomplished [26]. UVscope is a well-tested stand-alone NIST-calibrated multi-pixels photon detector, remotely controlled, designed to measure, in single photon counting mode, light flux with spectral acceptance matching the ones of the ASTRI camera [27]. Placed onto the primary mirror support and co-axial to the telescope receiving the light from the Illuminator, the UVscope will monitor the flux reaching the ASTRI aperture; this allows to completely eliminate the uncertainty due to the atmospheric transmission and propagation, so reaching an absolute telescope calibration. Among the several tests performed during the telescope and camera commissioning phases, two of them have been and are very important.

The first test demonstrated the viability of the ASTRI dual mirror SC design. On 24 May 2015, during the telescope structure and mirrors commissioning phase, a temporary CCD camera (~1° FoV) was installed at the focal plane of the telescope and the optical image of the North Star Polaris was taken [4]; this was the first ever optical image of the sky taken with a SC telescope. On October 2016, the Polaris has been newly observed with the temporary CCD camera under different offsets (from 0° to 4.5°) from the optical axis of the telescope demonstrating a constant optical PSF of a few arcmin over the full wide ASTRI SST-2M field of view [28, 29]. Furthermore, thanks to the analysis of several images acquired with the CCD camera till last Spring, the optical PSF shows an impressive stability.

The second enthusiastic result, that can be considered the first end-to-end test of the ASTRI SST-2M prototype, comes from the early commissioning phase of the ASTRI camera installed at the telescope focal plane. On 25 May 2017, during engineering test and first run on sky (dark moon), the first images of cosmic-ray events have been taken demonstrating the correct functioning of various camera components, in particular the ASTRI topological trigger (each event on the triggered PDM meets the trigger condition imposed), the CITIROC response (none discontinuity in the images dynamical range), the peak-detector operation mode (coherent signals across the PDMs containing the event). In the ASTRI SST-2M camera, both sensors and electronics are then successfully performing as designed [30]. Furthermore, all acquired data correctly passed to the camera server.

The first runs with the ASTRI SST-2M camera have been performed using an optical transparent Poly Methyl Metha Acrylate window to protect the camera from sand and dust and to keep a thermalized environment around the focal plane. It is under test a different material for the protection window, namely Spectrosil with multilayer interferential filters that, working well at different incident angles, reduces by a factor of 2 the detection of the night sky background.

Data produced during operation are stored in the ASTRI archive and are going to be analyzed by *A-SciSoft*, the official ASTRI data scientific analysis system, designed to comply with the general CTA requirements, to which a separate contribution is dedicated in these proceedings [31]. The software is able to analyze both real and Monte Carlo events from raw data to final scientific products and it has been successfully tested on simulated data, including a simulated, realistic observation of the Crab with the ASTRI SST-2M telescope prototype.

After completion of the overall ASTRI SST-2M commissioning in early Autumn, including the final tests of the data archive and analysis, the absolute calibration and science verification phases will follow with the purpose of assessing the instrument performance by means of observations targeted at bright TeV sources such as the Crab Nebula, Mrk 421 and Mrk 501. This science verification phase will cross-check the prototype performance with the predictions of detailed Monte Carlo simulations; we expect that an integral flux (E > 2 TeV) level of 1 Crab will be detected at 5σ significance in a few hours.

## 2.2 The ASTRI telescopes towards the construction of CTA

All components of the ASTRI SST-2M telescope prototype have been designed to comply with all CTA-SST requirements and are easy to be integrated in an array configuration such that foreseen for the ASTRI telescopes proposed to be installed at the CTA southern site. Nevertheless, some improvements in the ASTRI telescopes are foreseen, thanks to the experience acquired with the prototype. For example, the telescope structure will be lightened reducing the number of the mast structures and maintaining the same stiffness, while the coating of the mirrors will be optimized even in function of the material used for the camera protection window. New SiPM sensors will be adopted (in the prototype, the LCT5 model by Hamamatsu is used) and the capabilities of the CITIROC in the camera electronics will be upgraded to take into account the time tag of all triggered pixels. In any case, the ASTRI telescopes for the CTA will follow the design philosophy described in the previous paragraphs. Same consideration for what concerns the camera servers whose prototype has been already designed in view of the array configuration. In this case, the camera servers (one for each telescope), in addition to acquiring data without loss as for the prototype, have to interface with the CTA trigger, clock distribution, array control and data acquisition systems [12]. Furthermore, the ICT architecture designed for the prototype might be scaled up, taking into account the necessary control, monitor and alarm system requirements [16]. The Mini Array Software System (MASS) devoted to operate and control the ASTRI telescopes has been defined and proposed to CTA, developed upon the software frameworks chosen for the whole of CTA [32]. Eventually, data model, analysis pipeline and archive developed for the prototype are already designed in view of the array configuration [31].

The tenders process for the first nine ASTRI telescopes for the CTA is already started aiming to complete it by end of 2017. The installation of the first ASTRI telescope at the CTA southern site is foreseen in 2019 together with the external equipment devoted to its absolute calibration (this calibration equipment will be used even for every CTA telescope [33]).

The ASTRI telescopes for the CTA southern site will be located with inter-distance of the order of 250-300 m and they will able to verify the wide field of view performance and to investigate sources emitting at energies from a few TeV up to hundreds of TeV with energy resolution of the order of 10-15% [34]. The use of SiPMs improve the duty cycle because allows the extension of observations into periods with appreciable moonlight, as SiPMs do not suffer damage at high light levels. This increases the small size telescopes sensitivity where observations are limited by the number of signal gamma-rays observed rather than background levels. The ASTRI telescopes for the CTA southern site will reach a better sensitivity at energy greater than 10 TeV for extended sources (such as Supernovae Remnants and Pulsar-Wind Nebulae) with respect to current IACT experiments. Such a sensitivity could allow us to investigate the very-high-energy emission in different regions searching for possible spectral cut-offs [35]. Deep observation of a few selected targets will be performed in the high end of the gamma-ray energy range, even beyond 100 TeV, an almost unexplored energy region by IACTs [34]. The ASTRI project is part of the CTA and synergies with different CTA telescopes are important even during the early phases of the CTA observatory. As an example, preliminary simulations with an initial CTA array configuration including both ASTRI and MST telescopes have demonstrate the capability to search for dark matter signatures and more [36, 37]. In addition, fruitful synergies with HAWC, surveying a stripe of the northern sky accessible to pointed observation from the CTA southern site, are foreseen [4].

## 3. Conclusions

With more than one hundred IACT telescopes located in the northern and southern hemispheres, CTA will be the world's largest and most sensitive very-high-energy gamma-ray observatory. The energy range between a few TeV and 300 TeV will be covered by a large number of SST telescopes spread out over several square kilometers in the CTA southern site. The ASTRI telescopes are one of the proposed SST implementations with double-mirror design. For the early stages of the CTA southern site implementation, a set of nine ASTRI telescopes has been proposed. They constitute the main component of the ASTRI project, led by the Italian National Institute for Astrophysics, INAF, within the CTA context and with the collaboration of a number of Italian universities, the Universidade de Sao Paulo (Brazil), the North-West University (South Africa) and the Italian National Institute for Nuclear Physics, INFN.

The ASTRI SST-2M telescope installed in Italy, prototype of the ASTRI telescopes proposed for CTA, has already successfully confirmed the expectations related to its technological solutions: the optical system based on a dual-mirror Schwarzschild-Couder design with a curved focal plane covered by SiPMs sensors managed by a fast front-end electronics operating in peak-detector mode. The science verification phase will start early Autumn 2017.

The development of the ASTRI telescopes for the CTA is already ongoing aiming to install the first telescope at the CTA southern site in 2019, followed by the deployment of further (at least) eight ASTRI telescopes. This configuration will be able to verify the wide field of view performance and to investigate sources even beyond 100 TeV. Final aim of the ASTRI project is the construction of at least 35 out of the 70 small-sized telescopes envisaged for the CTA Southern site.

**Acknowledgments**

This work was conducted in the context of the CTA ASTRI Project. This work is supported by the Italian Ministry of Education, University, and Research (MIUR) with funds specifically assigned to the Italian National Institute of Astrophysics (INAF) for the Cherenkov Telescope Array (CTA), and by the Italian Ministry of Economic Development (MISE) within the "Astronomia Industriale" program. We acknowledge support from the Brazilian Funding Agency FAPESP (Grant 2013/10559-5) and from the South African Department of Science and Technology through Funding Agreement 0227/2014 for the South African Gamma-Ray Astronomy Programme. We gratefully acknowledge financial support from the agencies and organizations listed here: http://www.cta-observatory.org/consortium_acknowledgments.